\documentclass{article}
\usepackage[preprint]{neurips_2023}
\usepackage{booktabs}
\usepackage{array}
\usepackage{caption}
\usepackage{graphicx}
\usepackage{pgfplots}
\usepackage{ulem}  % Add this to your preamble

\usepackage[utf8]{inputenc} % allow utf-8 input
\usepackage[T1]{fontenc}    % use 8-bit T1 fonts
\usepackage{hyperref}       % hyperlinks
\usepackage{url}            % simple URL typesetting
\usepackage{booktabs}       % professional-quality tables
\usepackage{amsfonts}       % blackboard math symbols
\usepackage{microtype}      % microtypography
\usepackage{xcolor}         % colors
\usepackage{listings}
\usepackage{xcolor}
\usepackage{wrapfig} % For placing the figure beside the text

\usepackage{amsmath,amsfonts,bm}

\def\eqref#1{equation~\ref{#1}}
\def\1{\bm{1}}

\DeclareMathAlphabet{\mathsfit}{\encodingdefault}{\sfdefault}{m}{sl}
\SetMathAlphabet{\mathsfit}{bold}{\encodingdefault}{\sfdefault}{bx}{n}

\usepackage{hyperref}
\usepackage{url}

\usepackage{hyperref}
\usepackage{multirow}
\usepackage{fancybox}
\usepackage{enumitem}
\usepackage{graphicx}
\usepackage{balance}
\usepackage{color}
\usepackage{float}
\usepackage{tabularx}
\usepackage{adjustbox}
\usepackage{xcolor}
\usepackage{array}
\usepackage{amsmath}
\usepackage{xspace}
\usepackage{url}
\usepackage{tikz}
\usepackage{caption}
\usepackage{pgfplots}
\usepackage{listings}
\usepackage{color}
\usepackage{graphicx}
\definecolor{dkgreen}{rgb}{0,0.6,0}
\definecolor{gray}{rgb}{0.5,0.5,0.5}
\definecolor{mauve}{rgb}{0.58,0,0.82}
\pgfplotsset{compat=1.16}
\usepackage{pgf-pie}
\usetikzlibrary{pgfplots.statistics,calc}
\usepackage{algorithm}
\usepackage{algpseudocode}
\usepackage{subcaption}
\usepackage{listings}

\usepackage{tcolorbox}
\makeatletter
\newcommand{\mybox}[1]{%
	\setbox0=\hbox{#1}%
	\setlength{\@tempdima}{\dimexpr\wd0+13pt}%
	\begin{tcolorbox}[boxrule=0.5pt, colback=white, arc=4pt,
		left=6pt,right=6pt,top=6pt,bottom=6pt,boxsep=0pt]
		#1
	\end{tcolorbox}
}

\definecolor{codegreen}{rgb}{0,0.6,0}
\definecolor{codegray}{rgb}{0.5,0.5,0.5}
\definecolor{codepurple}{rgb}{0.58,0,0.82}
\definecolor{backcolour}{rgb}{0.95,0.95,0.92}

\lstdefinestyle{mystyle}{
  language=Python,
  aboveskip=3mm,
  showstringspaces=false,
  columns=flexible,
  numbers=none,
  backgroundcolor=\color{backcolour},
  commentstyle=\color{codegreen},
 keywordstyle=\color{magenta},
    numberstyle=\tiny\color{codegray},
    stringstyle=\color{codepurple},
    basicstyle=\small\ttfamily,
    breakatwhitespace=false,         
    breaklines=false,                 
    captionpos=b,                    
    keepspaces=false,                 
    numbersep=5pt,                  
    showspaces=false,                
    showstringspaces=false,
    showtabs=false,                  
    tabsize=2,
    escapeinside=``
}
\definecolor{nima2}{RGB}{1.0, 0.49, 0.0}
\definecolor{songcolor}{RGB}{191,191,191}
\definecolor{reemcolor}{RGB}{0.9, 5.0, 0.8}
\definecolor{aruncolor}{RGB}{51,255,51}

\title{SWE-Bench+: Enhanced Coding Benchmark for LLMs}

\author{%
Reem Aleithan, Haoran Xue, Mohammad Mahdi Mohajer, Elijah Nnorom, Gias Uddin, Song Wang \\ 
Lassonde School of Engineering \\
York University\\
\texttt{\{reem1100,hrx00,mmm98,ennorom,guddin,wangsong\}@yorku.ca} \\
}

\begin{document}

\maketitle

\begin{abstract}
Large Language Models (LLMs) in Software Engineering (SE) can offer assistance for coding. To facilitate a rigorous evaluation of LLMs in practical coding contexts, Carlos et al. introduced the \textit{SWE-bench} dataset, which comprises 2,294 real-world GitHub issues and their corresponding pull requests, collected from 12 widely used Python repositories. Several impressive LLM-based toolkits recently are developed and evaluated on this dataset. However, a systematic evaluation of the quality of SWE-bench remains missing. In this paper, we addressed this gap by presenting an empirical analysis of the \textit{SWE-bench} dataset. We conducted a manual screening of instances where \textit{SWE-Agent + GPT-4} successfully resolved issues 
%This is \sout{alongside a comparable sample of failed cases,} 
by comparing the model-generated patches with the actual pull requests. SWE-Agent+GPT-4 was at the top of SWE-bench leaderboard during the time of our study. 
%Through this approach, we aim to assess the robustness of LLMs in resolving real-world GitHub issues and uncover patterns in successful patches.  %\sout{that differentiate successful patches from failed attempts.}
% \song{describe which tools you used in your experiment, and how you analyze their data, 1. check the potential data leakage; 2. rigorously analyze the successful cases that can be fixed by LLMs to verify their robustness.}
Our analysis reveals some critical issues with the \textit{SWE-bench} dataset:
1) 32.67\% of the successful patches involve ``cheating'' as the solutions were directly provided in the issue report or the comments. We refer to as `solution leakage' problem. 2) 31.08\% of the passed patches are suspicious patches due to weak test cases, i.e., the tests were not adequate to verify the correctness of a patch. When we filtered out these problematic issues, the resolution rate of SWE-Agent+GPT-4 drops from 12.47\% to 3.97\%. 
% \song{need to update the number after tomorrow's manually check}.
{We also observed that the same data quality issues also exist in the two variants of SWE-bench, i.e., \textit{SWE-bench Lite} and \textit{SWE-Bench Verified}.} 
In addition, over 94\% of the issues were created before LLM’s knowledge cutoff dates, posing potential data leakage issues.
% \song{numbers need to be updated} 

 % \song{give some details/numbers regarding 1 and 2 from your manual analysis results.} 
 
The critical problem in the current versions of \textit{SWE-bench} dataset motivated us to refine it to build a more rigorous evaluation dataset \textit{SWE-Bench+}. We created SWE-bench+ by collecting GitHub issues that were created after the training cutoff dates of the LLMs to prevent the potential data leakage problem. 
%We performed the same data collection, filtering, patch-generation, and evaluation techniques described in the \textit{SWE-Bench} study. However, we focused on recent data where the issues and their corresponding pull requests were created after LLMs' training cutoff dates to avoid data leakage biases. 
We also ensure that the issues collected do not contain solutions in their reports or comments.  
% \song{also mention filter out issues that contain answers in the report}.  
% Furthermore, we analyzed the evaluation results and found that {solution-leak related issues are now resolved. However,} a prominent issue with weak test cases persists. 
After carefully analyzing the passed instances from the \textit{SWE-Agent + GPT-4} model with the new dataset, \textit{SWE-Bench+}, we observed a decline in the pass rate, dropping from 3.97\% (as seen on the refined \textit{SWE-Bench}) to a resolution rate of 0.55\%. 
We further evaluated \textit{SWE-RAG + GPT-4}, \textit{SWE-RAG + GPT-3.5},  and \textit{AutoCodeRover + GPT-4o} models on the new dataset to verify our findings, where the resolution rates of the models drop significantly, which are 0.73\%, 0.55\%, and 3.83\%, respectively. 

% \song{describe which tools you experiment with and how much the performance changed}

\end{abstract}

\section{Introduction}
\label{sec:intro}
%\song{https://arxiv.org/pdf/2305.01210}
%\song{https://www.eecs.yorku.ca/~wangsong/papers/fse22b.pdf}
%\song{story line: We conduct the first empirical study to explore the quality of SWE-BENCH dataset, we find some comment issues 1) xx\% issues and xxx\% patches of the isssues were created before LLM's knowledge cutoff dates, i.e., the dates at which training data were gathered.
%%2) suspicious patches caused by weak test cases;
%}
\textit{SWE-bench} aka Software Engineering Benchmark dataset is created to systematically evaluate the capabilities of an LLM in resolving software issues. The dataset contains 2,294 complex issues from GitHub~\cite{jimenez2024swebenchlanguagemodelsresolve}. Given as input the issue information to an LLM, the task for the LLM is to modify the code base to address the issue (i.e., resolution). Each input for an issue consists of a description and a pull request with a reference to the corresponding buggy code repository. 
Each issue can be either a bug report or a new feature request. The pull request contains the code changes made by developers to address the issue, along with test cases designed to check if the feature is properly implemented or if the bug is successfully fixed. Two variants of the \textit{SWE-bench} datasets are recently developed: \textit{SWE-bench Lite}\footnote{https://www.swebench.com/lite.html} and \textit{SWE-bench Verified}\footnote{https://openai.com/index/introducing-swe-bench-verified/}. 
\textit{SWE-bench Lite} focuses on 300 issues related to bug fixing. \textit{SWE-bench Verified} contains 500 verified issues with clear issue descriptions and strong test cases.

A significant body of work from both academia and industry has so far utilized \textit{SWE-bench} and its variants to develop and to test LLM coding capabilities~\cite{chen2024coderissueresolvingmultiagent,zhang2024diversityempowersintelligenceintegrating, xia2024agentlessdemystifyingllmbasedsoftware, yang2024sweagentagentcomputerinterfacesenable, zhang2024autocoderoverautonomousprogramimprovement, larosa2024githubissuessolvedtree, zan2024swebenchjavagithubissueresolving}. 
Given an issue and its associated buggy code repository, these LLM-based approaches can perform a series of complex tasks, such as reasoning about the target bug's location, analyzing the root cause of the issue, proposing strategies for fixing the bug, and ultimately writing a patch to fix the issue.  
Within less than one year, the resolution rate on \textit{SWE-bench Full} increased from 0.17\% (for RAG+GPT3.5) to around 22.00\% (for Honeycomb). The performance of the LLMs on \textit{SWE-bench Lite} and \textit{Verified} went up to 45\%. 

However, \textbf{\textit{are the LLMs actually resolving the issues in SWE-bench?}} 

In this paper, we answer the above question by offering two contributions. \textbf{First,} we present an empirical study of state-of-the-art (SOTA) LLMs on SWE-bench Full that explores 1) the quality of SWE-bench issues with a focus on the testing adequacy of the test cases used for validating patches and 2) the quality of patches generated by the LLMs to fix the issues. \textbf{Second,} we present an enhancement of \textit{SWE-bench Full}, which we call \textit{SWE-bench+}.

During our empirical study, \textit{SWE-Agent+GPT-4} was at the top of the \textit{SWE-bench} online leaderboard. Other top approaches (e.g., Honeycomb, Amazon Q Developer Agent, and Factory Code Droid) were either closed-sourced commercial tools or not verified by the SWE-bench team regarding reproducibility. SWE-agent~\cite{yang2024sweagentagentcomputerinterfacesenable} allows LLM agents to execute basic file operations via shell commands to achieve interaction between the LLM engine and a software repository. First, we picked issues claimed as resolved by \textit{SWE-Agent+GPT-4}. We did this by filtering only the instances with evaluation logs showing that all tests passed. Second, we performed a patch validation study by comparing the gold patches (i.e., original) to the model patches (i.e., generated). We did this by comparing the files changed, the lines changed, and the code changes made in the fixes (both original and generated). Third, we determined eight patterns in the generated fixes by reviewing the issue reports, the corresponding tests, and the available discussions of issues (which are treated as hints to LLMs). 

\begin{figure}[t]
    \centering
    \begin{subfigure}[b]{0.48\textwidth}
        \centering
\includegraphics[width=\textwidth]{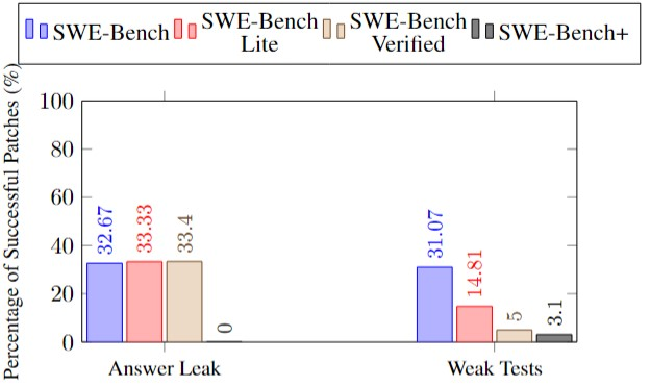}
        \caption{Answer Leak vs Weak Tests across Datasets}
        \label{fig:answer-leak-weak-tests}
    \end{subfigure}
    \hfill
    \begin{subfigure}[b]{0.48\textwidth}
        \centering
%\resizebox{\textwidth}{!}
\includegraphics[width=\textwidth]{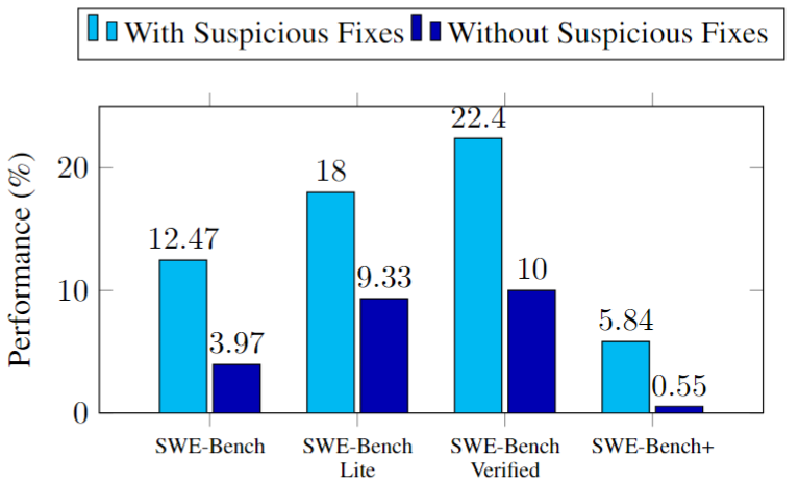}
        \caption{Performance of SWE-Agent + GPT-4} 
        \label{perf-dist}
    \end{subfigure}
    \caption{Comparison of performance metrics and patterns across \textit{SWE-bench} datasets}
    \label{fig:combined-results}
\end{figure}

%\song{you should say something negative here, these are patterns that affect the quality of SWE-bench, not just "patterns"}:
Our identified six patterns in the 251 SWE-Agent+GPT-4 patches can be broadly divided into two types: \textit{suspicious fixes} and \textit{correct fixes}. Suspicious fixes corresponded to 63.75\% (i.e., 160) of the patches. Two patterns were prevalent in those fixes: 
% \gias{these patterns do not match with Table \ref{pattern-analysis-table}}.  
 %\gias{OK, but that does not prove that there is a data leak. Even a claim for potential data leak is a bit too strong. Let's justify it this way: ``a major source of code related training data in LLMs comes from the GitHub open source software (OSS) repositories. Trivially, all OSS repos are used for training, except those explicitly prohibit to be included. This means that potentially all issue reports in the SWE-bench full dataset may be already seen by the LLMs during their training phases, given the issue reports in the SWE-bench full dataset were taken during/before the training of the LLMs. While, without exclusive access to the training data itself it is not possible to confirm of an overlap between the SWE-bench full dataset and the training data of LLMs, the overlap in their time of data collection does raise the concern of potential data leak."}. }  
1) \textbf{Answer Leak.} In 32.67\% of the resolved instances, the solutions were outlined directly in the issue reports or comments. %This brings into question the model's true performance and the reliability of SWE-Bench as a benchmark. When the model relies on pre-existing solutions, it doesn’t showcase real problem-solving skills, limiting its ability to be evaluated on generating original solutions.
% 2) \textbf{Potential Data Leak.} 94\% of the instances in SWE-bench and their pull requests were created before the models'\song{which model} training cut-off dates posing potential data leak-related issues.
%In fact, we found that some of the passed patches were either identical to the gold patches or showed only minor differences in style and implementation. 
%\song{numbering the patterns} 
2) \textbf{Weak Tests.} In 31.08\% of the resolved instances, the changes made by the model are either incorrect, incomplete, or applied to different files or functions compared to the gold patch. Despite these discrepancies, the changes pass the tests, indicating that the tests are too weak to catch such errors.  In addition, the dataset can also suffer from potential data leak issues. This is because 94\% of the instances in SWE-bench and their pull requests were created prior to the training cut-off dates of the LLMs, meaning that all issue reports in the full SWE-bench dataset may have been exposed to the LLMs during their training phases, raising concerns about potential data leakage.

% \song{rewrite this, the weak tests make these simplistic patches pass}Some patches were overly simplistic, missing critical scenarios, while others included errors or irrelevant changes in unrelated files or functions.

%Only a small percentage of the patches were original, newly crafted solutions that correctly addressed the issue, with an even smaller fraction offering more comprehensive fixes. 

Based on the above observation, we considered fixes corresponding to issues with answer leakage and weak tests as `suspicious fixes'. In Figure \ref{fig:answer-leak-weak-tests}, we show the distribution of such fixes in the three literature datasets (i.e., SWE-bench full, lite, verified). In Figure \ref{perf-dist}, we show that after filtering these suspicious fixes, the correct resolution rate of \textit{SWE-Agent+GPT-4} dropped to 3.97\% from 12.47\%. This drastic drop in resolution rate raises concerns about the robustness of the model-generated patches and the reliability of the SWE-bench dataset itself. 

To address the problems in \textit{SWE-bnech} datasets, we created \textit{SWE-bench+} dataset, which ensures that: 1) the data were created after the models' training cut-off dates, and 2) the issues do not include solutions in the issue description or comments. \textit{SWE-bench+} dataset is created by following the same data collection methodology described in the \textit{SWE-Bench} dataset; except we filtered out issues with answer leakage problems. Considering the training cut-off dates of the LLMs used in our study—GPT-3.5 (turbo-16k-0613) with a cut-off in September 2021, GPT-4 (1106) up to April 2023, and GPT-4o (2024-05-13) up to October 2023, we opted to collect data starting a month after the most recent model's cut-off date. To the end, we gathered issues from the period of 2023-11-01 to 2024-08-22. As we show in Figure \ref{fig:answer-leak-weak-tests}, \textit{SWE-bench+} has no issues with solution leakage whereas all the other three datasets suffer from this. \textit{SWE-bench+} also has the lowest proportion of issues with weak test cases among all the SWE-bench variants. As such, we consider SWE-bench+ as the most robust dataset among the available \textit{SWE-bench} variants. When we ran \textit{SWE-Agent+GPT-4} on \textit{SWE-bench+} dataset, its resolution rate dropped to 0.55\% (see Figure \ref{fig:combined-results}). 

We further evaluated the \textit{SWE-RAG + GPT-4}, \textit{SWE-RAG + GPT-3.5}, and \textit{AutoCodeRover + GPT-4o} models on the new dataset to verify our findings. The resolution rates of the models dropped significantly, with the new rates being 0.73\%, 0.55\%, and 3.83\%, respectively. In comparison, the previously reported resolution rates on the SWE-Bench leaderboard were 1.31\% for \textit{SWE-RAG + GPT-4}, 0.17\% for \textit{SWE-RAG + GPT-3.5}, and 18.83\% for \textit{AutoCodeRover + GPT-4o}.

% \gias{drop from where? we don't talk about their original resolution rate here}.
% \reem{need to include original rates from leaderboard}

\noindent \textbf{Artifacts.} While working on merging our \textit{SWE-bench+} to \textit{SWE-bench} project repository, we release the dataset of \textit{SWE-bench+} to help other researchers replicate and extend our study\footnote{\url{https://zenodo.org/records/13879453}}.%\song{create a https://zenodo.org/ repo and share our data}

\begin{figure}[t!]
    \centering
    \includegraphics[width=0.9\textwidth]{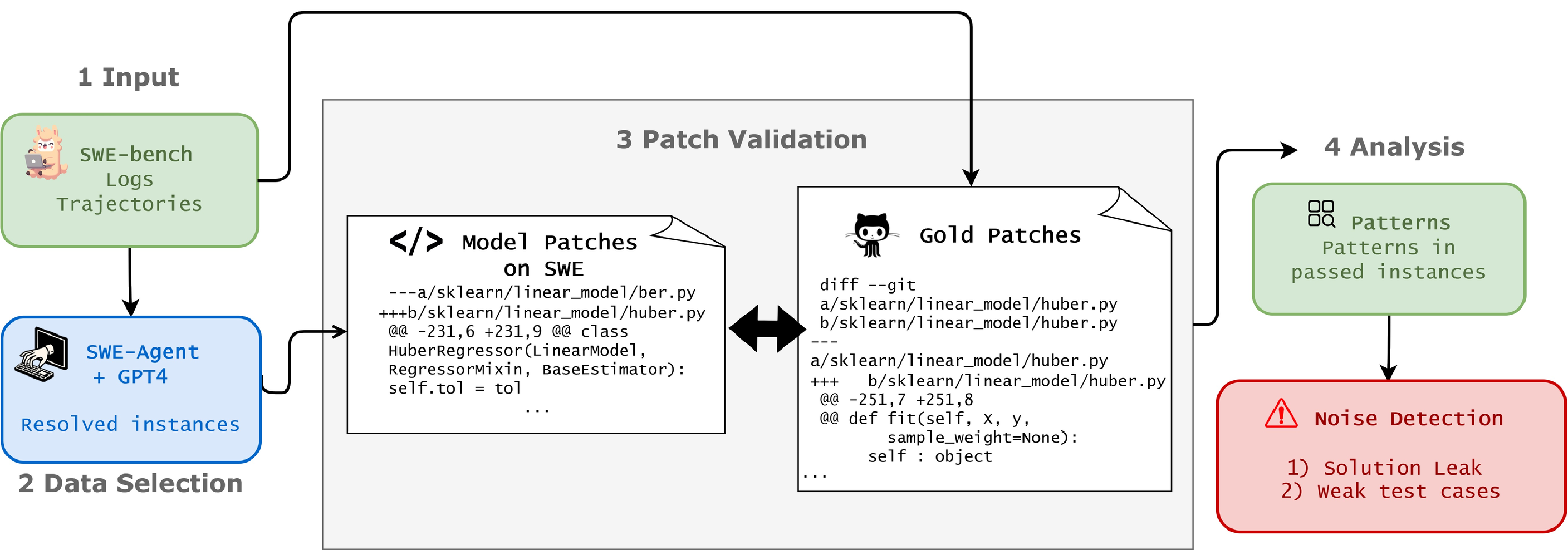} % Adjust width as needed
    \caption{Overview of robustness analysis for \textit{SWE-Bench} datasets}
    \label{swe-image}
\end{figure}

%\subsection{Empirical Study}
\section{Robustness Analysis of SWE-Bench}
\label{sec:2}

% \song{Please update the patterns, we have new ones}

%Figure \ref{swe-image} shows the overflow of our empirical study. %how we empirically evaluated the performance of \textit{SWE-agent + GPT-4} on \textit{SWE-bench} by performing a patch-validation study over the model-generated patches and the gold patches. We also explain how we analyzed the average lines changed to understand the complexity of the resolved instances as compared to the failed instances.

%\subsection{Empirical Study}
We conducted an empirical study of \textit{SWE-Agent+GPT-4} generated patches for issues in the \textit{SWE-bench Full} dataset. The goal of our study was to identify whether the patches exhibit any potential problems.  %and if so determine the \textit{types} of the problems. We refer to each such type a pattern.  
Figure \ref{swe-image} outlines the major steps we followed in our study. The input is the set of all issues in \textit{SWE-bench}. Each issue contains a description and the patch to address the issue. Each patch is a diff of code changes. We call this a ``gold patch''. We picked \textit{SWE-Agent + GPT-4} and applied it on each issue to create a fix. We refer to the model output for an issue as a ``generated patch''. We then compared the gold and generated patches to an issue by analyzing the corresponding files changed in the pull requests on GitHub with the same \textit{instance\_id}. As we studied the model generated patches, we also examined the logs and trajectories generated by the model. Logs provide the step-by-step execution of the models. The trajectory data provide a detailed record of the models’ decision-making processes while making a resolution as a patch.%, showing how they iterated through possible solutions before arriving at a final patch. 

% The comparison between the gold and generated patches was essential to evaluate how closely the model's output aligned with the developer-generated patches and to assess the quality of the solutions provided. 
To reduce potential biases during the comparison between gold and generated patches, three authors independently performed the patch validation study. Each author carefully examined the files and lines changed, reviewed the issue descriptions, and evaluated the implementation styles and intentions behind both the model-generated and developer-generated patches. {The disagreements were resolved through a broader discussion involving all the authors.} 

For the model generated patches, we focused on instances where the generated patches resolved the issue and passed all associated tests. As a result, we identified 251 instances from the \textit{SWE-Bench Full} dataset. Note that, to ensure the patches passed all tests, we reviewed the evaluation logs of 286 instances initially marked as resolved by \textit{SWE-Agent + GPT-4} in the \textit{results.json} file from the \textit{SWE-Bench Full} evaluation repository and selected only those with logs confirming that all tests passed following the application of the generated patches.

% \textbf{Patch-Validation Study}.  
% Next, as seen in Figure \ref{swe-image}, we collected the model-generated patches from the \textit{.diff} files located in the \textit{experiment} directory for \textit{SWE-Agent+GPT-4}. 

\subsection{Critical issues of \textit{SWE-Bench}}
%\song{also needs to describe how each category is created.}
Among the 251 generated patches that passed all test cases in \textit{SWE-bench Full} dataset, we found several patches as problematic/suspicious. Table \ref{pattern-analysis-table} outlines six patterns in the 251 generated patches, four related to the suspicious fixes and two related to the correct fixes. To explain each pattern, we provide definitions, including the number of instances associated with each pattern and the likely root causes. We discuss each pattern below. 
%Further detailed examples for each pattern can be found in \ref{patterns_sec}. 
\begin{table}[t!]
\caption{Patterns found among the 251 successful patches generated by \textit{SWE-Agent + GPT-4}}

\label{pattern-analysis-table}
\begin{center}
\resizebox{0.9\textwidth}{!}{
\begin{tabular}{llcc}
\toprule
{\bf Type} & {\bf Pattern} & {\bf Numbers (percentage)} & {\bf Root cause}  \\  \midrule

\multirow{5}{*}{Suspicious fixes} & Solution leak & 82 (32.67\%) & solution leakage \\ %\cmidrule{2-4} 
& Incorrect fixes  & 32 (12.75\%) & weak tests \\ %\cmidrule{2-4} 
 
& Different files/functions changed & 9 (3.59\%) & weak tests \\ %\cmidrule{2-4}
& Incomplete fixes & 37 (14.74\%) & weak tests \\ \midrule
\multirow{3}{*}{Correct fixes}  
& {Different fixes from gold patches} & 76 (30.27\%) & --  \\ %\cmidrule{2-4} 
& {More comprehensive fixes than gold patches} & 15 (5.98\%) & -- \\
\bottomrule
\end{tabular}
}
\end{center}
\end{table}

%\todo{show an example for each category}

\subsubsection{Patterns extracted from the suspicious fixes}
\label{patterns_sec}
We observed four patterns in the suspicious fixes, one is attributed to solution leakage and the other three are attributed to weak test case problems in the \textit{SWE-bench} dataset.

\begin{figure}[t!]  % Use [H] to enforce exact placement
    \centering
    \includegraphics[width=0.9\textwidth]{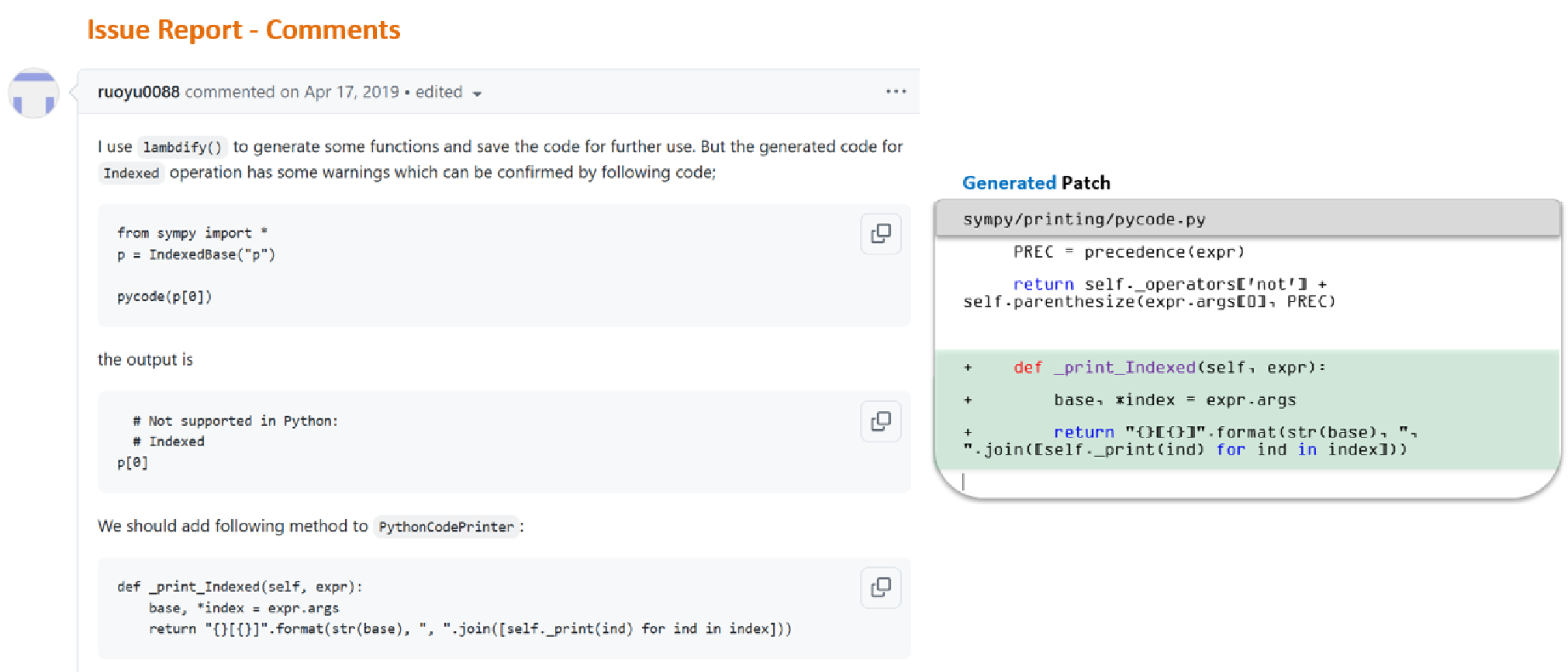} % Adjust the path to the image
    \caption{Solution Leakage in issue report for sympy-16669}
    \label{sol_leak-image-3}
\end{figure}

% \song{explain every example, from which project, what are the reasons;}
%\begin{enumerate}
\noindent \textbf{1. Solution leak:} represents instances where the solution to the issue is clearly outlined in the issue description or comments on GitHub. Since both the issue descriptions and comments (referred to as \textit{hints\_text} in the SWE-Bench study) are provided as input to the models, these LLM models can extract the solutions directly from this information instead of generating it independently. 
32.67\% of the successfully resolved issues followed this pattern, making it the most common among resolved patches. This raises significant concerns about a model's actual performance and the validity of the SWE-Bench instances as benchmarks. If a model is simply copying the solution it already has access to, it isn't demonstrating true problem-solving capabilities but rather replicating what is provided, thus limiting the assessment of its ability to generate new solutions.
The example shown in Figure \ref{sol_leak-image-3} illustrates issue report 16669\footnote{\url{https://github.com/sympy/sympy/issues/16669}} from the \textit{sympy} project, where the issue description provided the exact solution code patch required to resolve the issue, which makes it possible for the model to directly copy the solution from the issue report and generate the same solution as provided.

\begin{figure}[t!]
    \centering
    \includegraphics[width=0.8\textwidth]{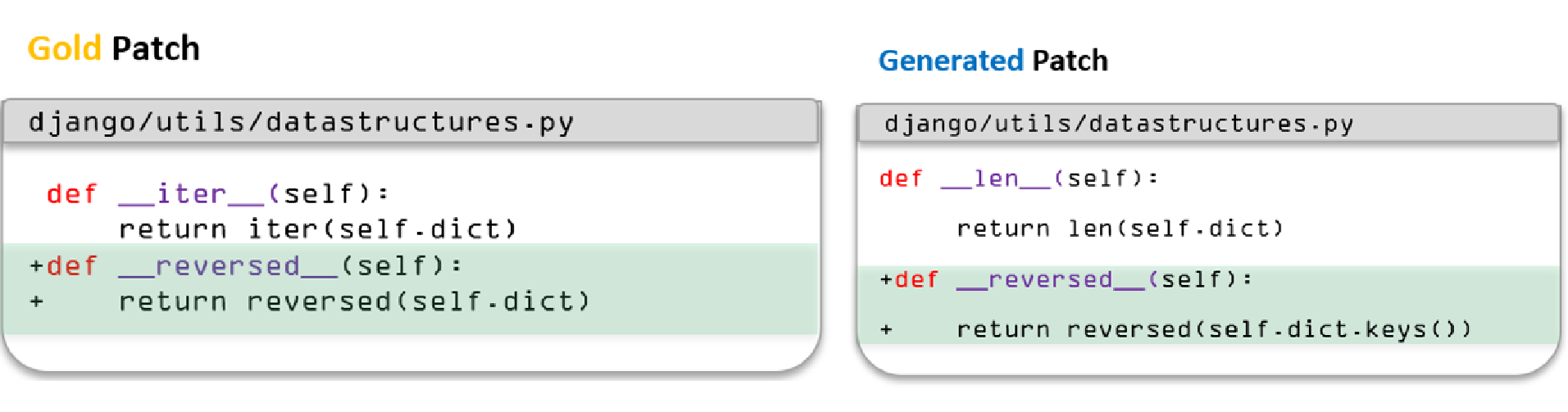} % Adjust the path to the image
    \caption{Incorrect fix generated by the model for django-32517}
    \vspace{-0.15in}
    % \song{give the issue id}
    \label{wrong-image}
\end{figure}

\noindent \textbf{2. Incorrect fixes:} refer to cases where the model-generated patches provide incorrect solutions, yet pass the test cases when they should have failed. 
This pattern was present 12.75\% of the passed instances. Suggesting a weakness in test cases where the functionality of the issue resolution is not correctly captured. 
The fact that incorrect patches can pass the test cases raises suspicion about the relevance and accuracy of the test cases in assessing whether the issue has been fully resolved. Figure \ref{wrong-image} shows a comparison between the model-generated patch and the gold patch for django-32517\footnote{\url{https://code.djangoproject.com/ticket/32517}}. According to the issue description, a new functionality is needed to reverse a Python \textit{OrderedSet} by implementing the \textit{\_\_reversed\_\_} function. The gold patch demonstrates the correct behavior, where the entire dictionary is reversed, while the generated patch only reverses the dictionary’s keys. As a result, the two patches produce entirely different outputs, as they apply different methods to the dictionary.

\begin{figure}[h]  % Use [H] to enforce exact placement
    \centering
    \includegraphics[width=0.85\textwidth]{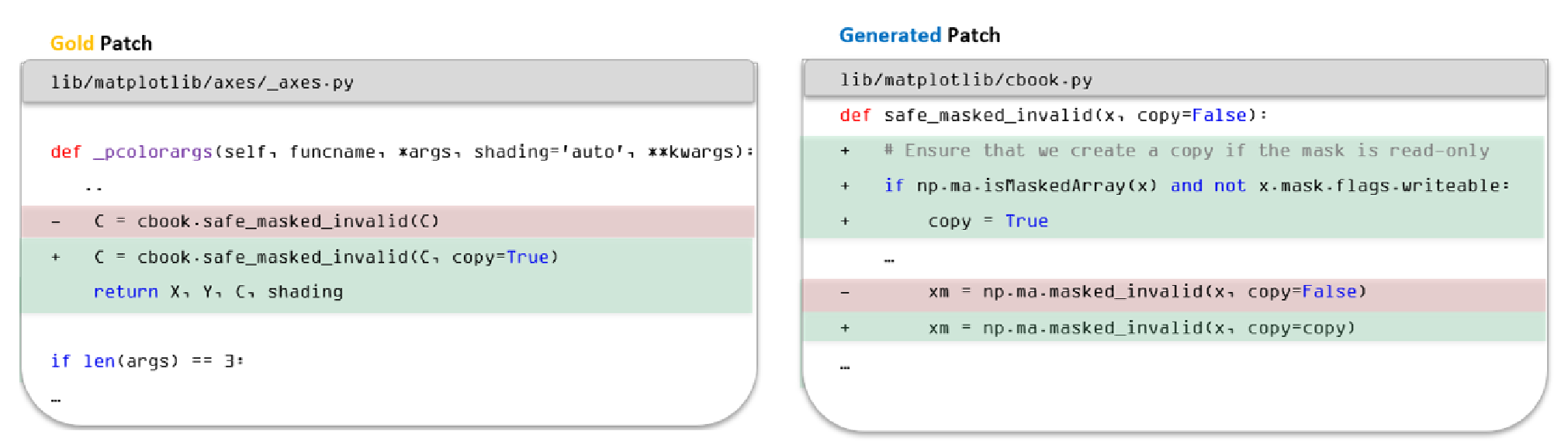} % Adjust the path to the image
    \caption{Different files changed by model for issue-26093 of Matplotlib}
    \label{diff_files-image}
\end{figure}

\noindent \textbf{3. Different files/functions changed:}
This pattern refers to cases where the model-generated patches modify files or functions unrelated to the issue at hand. These files differ from those altered in the gold patch, yet the model’s patches still pass the test cases despite this discrepancy. This highlights a weakness in the model’s ability to accurately locate and address the source of the issue. The fact that the test cases pass, even though changes were made in irrelevant files, suggests that the test cases are either weak or irrelevant and should have failed in detecting the incorrect modifications. Figure \ref{diff_files-image} presents an example from issue-26093 of Matplotlib project\footnote{\url{https://github.com/matplotlib/matplotlib/issues/26093}}, where the model-generated patch modifies the \textit{cbook.py} file, while the gold patch makes changes to the \textit{\_axes.py} file. This shows that the model's patch affects a completely different file from the gold patch, highlighting the model's inability to accurately identify the correct file containing the bug.

\noindent \textbf{4. Incomplete fixes:}  
This pattern refers to model-generated patches that offer incomplete implementations compared to the gold patches, often omitting critical details. For instance, some patches include only partial if-else statements, neglecting edge cases that the gold patch addresses. Although the model-generated patches follow the correct implementation approach, they overlook important aspects that could lead to failures in production or when handling edge cases. This underscores a weakness in the test cases, as they fail to capture the finer details necessary for a comprehensive issue resolution.

\begin{figure}[t!]  % Use [H] to enforce exact placement
    \centering
    \includegraphics[width=0.9\textwidth]{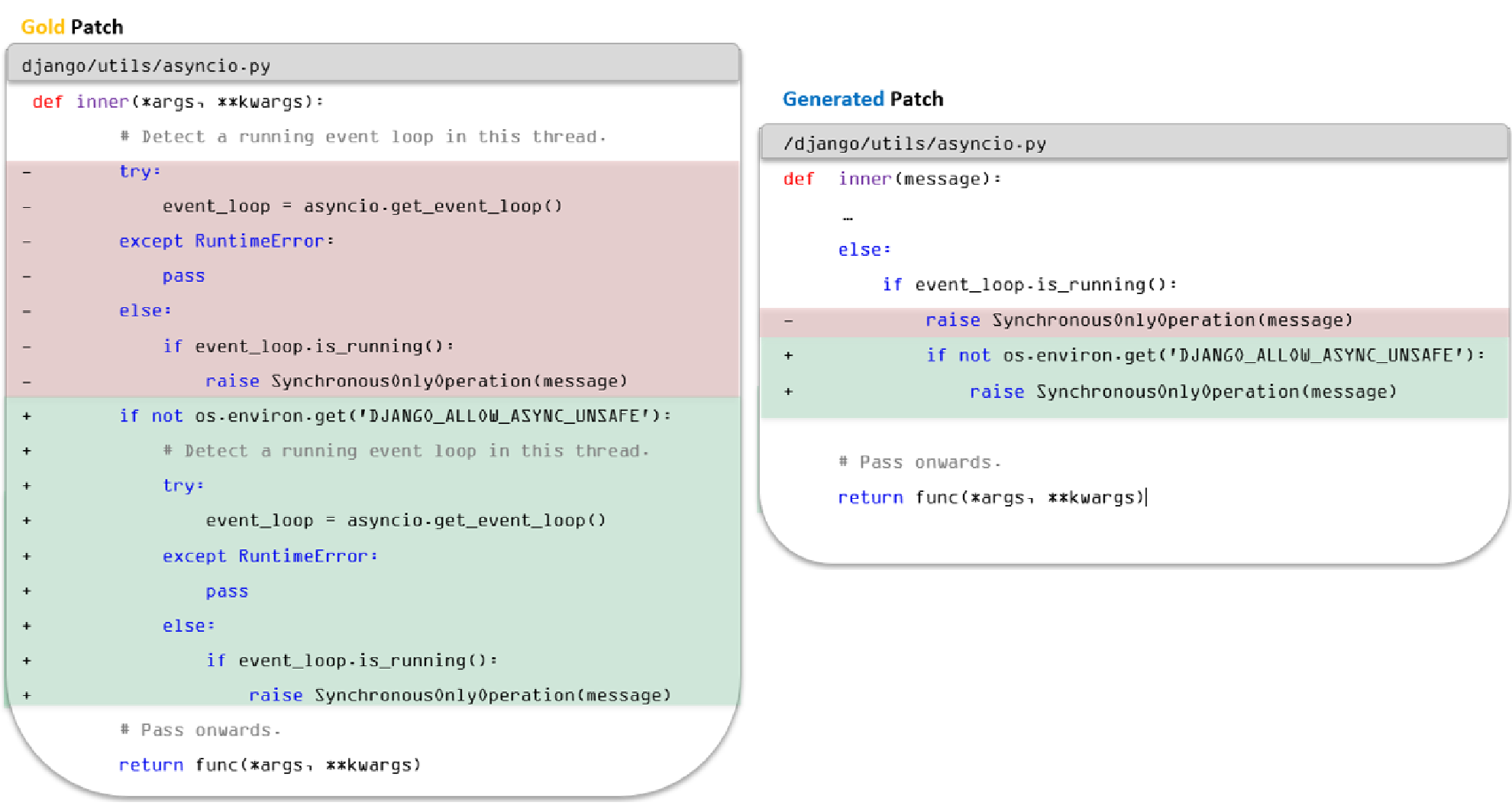} % Adjust the path to the image
    \caption{Incomplete fix generated by the model for django-31056}
    % \song{give the issue id}
    \label{incomplete-image}
\end{figure}

The example provided in Figure \ref{incomplete-image} shows the same change being made by the model and the one made by the developers in the gold patch\footnote{\url{https://code.djangoproject.com/ticket/31056}}. The gold patch provides a complete fix while the model patch provides a partial fix. 
Specifically, the gold patch properly handles the detection of an event loop in the current thread by including a \textit{try-except} block to catch \textit{RuntimeError} when an event loop is unavailable and checks if the event loop is running before raising an exception. Additionally, it wraps the entire logic in a condition that checks the environment variable \textit{DJANGO\_ALLOW\_ASYNC\_UNSAFE}. In contrast, the generated patch is missing critical parts of this logic, such as the \textit{try-except} block and the check for a running event loop. As a result, the model-generated patch is incomplete, missing key error handling and flow control that are necessary for ensuring safe operation.

% \song{give an example here}

\subsubsection{Patterns extracted from the correct fixes}
%\begin{enumerate}        
% \item \noindent \textbf{Exact the same as gold patches:}
% This pattern occurs when the model-generated patches are identical to the gold patches created by the developers. In some instances, this even applies to longer code patches. The similarity between the model-generated patch and the gold patch often extends to developer comments, variable names, and exact spacing. This pattern represents 9.14\% of the passed instances are the same as the gold patches. 

 \noindent \textbf{1. Different fixes from gold patches:}
This pattern refers to cases where the model-generated patches present an entirely different solution to the issue compared to the gold patch. Although the coding style and implementation differ, the model-generated patches correctly resolve the issue. Instances in this pattern are considered a correct resolution of the issues. 

\noindent \textbf{2. More comprehensive fixes:}
In contrast to the previous pattern, this refers to instances where the model-generated patches are more comprehensive than the gold patches. For example, the model-generated patches may include additional if-else cases or other logic that the gold patch does not address. 
This showcases a strength of LLMs, as they can generate more thorough solutions that cover scenarios developers might overlook, resulting in potentially safer and more robust solutions (e.g., by incorporating try-catch statements or handling edge cases more effectively). We consider this pattern as a correct resolution of the issue. 
\subsection{Updated Resolution Rate of SWE-Agent + GPT-4 on SWE-Bench Full}

After identifying these patterns, we recalculated the resolution rate of \textit{SWE-Agent + GPT-4}, focusing exclusively on patches classified under the correct fixes patterns. This new resolution criteria resulted in a significant drop in the resolution percentage, as shown in Figure \ref{perf-dist}. The performance of \textit{SWE-Agent+GPT-4} decreased from 12.47\% to 5.49\% when considering only correct fixes—those where the model generated either correct but different implementations from the gold patch (30.27\% of the time) or more comprehensive fixes than the gold patch (5.98\% of the time). We excluded all suspicious fixes, which included instances where the model generated incorrect solutions (12.75\%), where the issue report contained a direct solution (32.67\%), where changes were made in files unrelated to the gold patch (3.59\%), and cases where the model produced incomplete fixes, missing critical details of the solution (14.74\%).

% \song{give more detailed results numbers here}

\subsection{Updated Resolution Rate of SWE-Agent+GPT-4 on SWE-bench Lite and Verified}

% \song{revisit the suspicious numbers in table 1, and see how many are in these two now, create a table}

Two new variants of \textit{SWE-Bench} are recently developed, \textit{SWE-Bench Lite} and \textit{SWE-Bench Verified}, each designed with different goals. \textit{SWE-Bench Lite} focuses on instances with lower evaluation costs and increased accessibility. On the other hand, \textit{SWE-Bench Verified} aims to provide a curated subset of \textit{SWE-Bench}, where human annotators filter out issues with underspecified descriptions or weak unit tests that might reject valid solutions. However, neither of these new datasets addresses the solution leakage problem, which was the primary motivation for our study. To investigate this, three of the authors reviewed all issue reports for the instances in \textit{SWE-Bench Lite} and \textit{SWE-Bench Verified} for cases of solution leakage. \textit{In SWE-Bench Lite}, we identified 18 instances where the solution was directly provided in the issue description or discussion on GitHub. Similarly, in \textit{SWE-Bench Verified}, 37 instances contained direct solutions in either the issue description or the discussion on GitHub. Table~\ref{lite-verified-patterns} shows more details.

\begin{table}[t!]
\centering
\caption{Patterns found among SWE-Bench Lite (total passed: 54, 18.0\%) and SWE-Bench Verified (total passed: 112, 22.4\%) datasets with successful patches generated by SWE-Agent + GPT-4.}

\label{lite-verified-patterns}

\resizebox{0.95\textwidth}{!}{%
\begin{tabular}{lrrrr}
\toprule
\textbf{Pattern} & \textbf{SWE-Bench Lite} & \textbf{Percentage (Lite)} & \textbf{SWE-Bench Verified} & \textbf{Percentage (Verified)} \\ \midrule
\textbf{Incorrect fixes} & 5 & 9.26\% & 14 & 12.50\% \\ %\hline
\textbf{Incomplete fixes} & 3 & 5.56\% & 11 & 9.82\% \\ %\hline
\textbf{Different Files/Functions Changed} & 0 & 0.00\% & 0 & 0.00\% \\ %\hline
\textbf{Solution Leak} & 18 & 33.33\% & 37 & 33.04\% \\ \bottomrule
\end{tabular}%
}
\end{table}

\begin{comment}
\begin{table}[t!]
\centering
\caption{Patterns found among \textit{SWE-Bench Lite} and \textit{SWE-Bench Verified} with successful patches generated by \textit{SWE-Agent + GPT-4}.}
\label{lite-verified-patterns}
\vspace{-.01in}
\resizebox{0.95\textwidth}{!}{%
\begin{tabular}{l|c|c|c|c}
\hline
\textbf{Pattern} & \textbf{SWE-Bench Lite} & \textbf{Percentage (Lite)} & \textbf{SWE-Bench Verified} & \textbf{Percentage (Verified)} \\ \hline
\textbf{Incorrect} & 5 & 9.26\% & 14 & 12.50\% \\ \hline
\textbf{Incomplete} & 3 & 5.56\% & 11 & 9.82\% \\ \hline
\textbf{Different Files/Functions Changed} & 0 & 0.00\% & 0 & 0.00\% \\ \hline
\textbf{Answer Leak} & 18 & 33.33\% & 37 & 33.04\% \\ \hline
\end{tabular}%
}
\end{table}
\end{comment}
% \song{create a table to show the number of each type of examined suspicious issues that exist in these two variants}

% \reem{describe table}
We also observed additional suspicious fixes in both \textit{SWE-Bench Lite} and \textit{SWE-Bench Verified}, as summarized in Table \ref{lite-verified-patterns}. Specifically, 9.26\% of fixes in \textit{SWE-Bench Lite} and 12.50\% in \textit{SWE-Bench Verified} were incorrect fixes generated by \textit{SWE-Agent+GPT-4}. Additionally, 5.56\% of fixes in \textit{SWE-Bench Lite} and 9.82\% in \textit{SWE-Bench Verified} were incomplete fixes from the same model. Notably, there were no issues involving changes to different functions or files, as the model correctly identified the buggy file in all cases for both datasets. Despite this, the overall suspicious fix patterns led to 48.14\% suspicious fixes in \textit{SWE-Bench Lite} and 55.36\% in \textit{SWE-Bench Verified}, significantly reducing the resolution rates—from 18\% to 9.33\% in \textit{SWE-Bench Lite} and from 22.4\% to 10.0\% in \textit{SWE-Bench Verified}. 
% \song{here say  18 of lite had solution  in issue report and 37 of verified had solution in issue report}
  
%\subsection{Average number of lines changed (passed vs failed):}
%To better understand the nature of the issues successfully resolved versus those that failed, we calculated the average number of lines changed by SWE-Agent + GPT-4 for both the passed and failed instances. As shown in Table \ref{line-change-analysis-table}, the patches that passed involved an average of only 15.04 lines changed, while the failed patches averaged 48.52 lines. This suggests that the model is more effective at resolving simpler issues with fewer changes, while more complex issues, requiring a greater number of modifications, tend to fail.
%\begin{table}[H]
 %   \caption{Average Number of Lines Changed by SWE-Agent + GPT-4 Model}
 %   \label{line-change-analysis-table}
%    \begin{center}
  %  \begin{tabular}{lc}
 %   \hline
%    \multicolumn{1}{c}{\bf Status} & \multicolumn{1}{c}{\bf Average Number of Lines Changed} \\
%    \hline
 %   Passed & 15.04 \\
 %   Failed & 48.52 \\
%    \hline
%    \end{tabular}
%    \end{center}
%\end{table}

% \begin{figure}[h!]
%     \centering
%     \includegraphics[width=\textwidth]{swe+.png} % Adjust width as needed
%     \caption{SWE-Bench+ Methodology  \gias{explain this at the start of the section}}
%     \label{swe+-image}
% \end{figure}
% \song{aslo no need to have Figure 3, it's the same as SWE-Bench}

\section{Building SWE-Bench+}
% \song{these reasons need to be updated}
%As discussed in the previous section, \textit{SWE-Bench} was found to have solution leakage issues, which raised concerns about the robustness of the dataset. 
%In addition, 
%Our analysis reveals some critical issues with the SWE-bench dataset: 
%\add{1) over 94\% of the issues were created before LLM’s knowledge cutoff dates, posing potential data leakage issues.} 
%2) 32.67\% of the successful patches involve ``cheating'' as the solutions were directly provided in the issue report or the comments. This leads to a clear case of solution leakage, where the model relies on existing solutions rather than generating them independently, and 3) 31.08\% of the passed patches are suspicious patches due to weak test cases. As a result, the accuracy of the-used-tool drops from 12.47\% to 3.97\%. 
% \song{need to update the number after tomorrow's manually check}.
%{In addition, we also observed that the revealed data qualify issues also exist in the two variants of SWE-bench, i.e., SWE-bench Lite and SWE-Bench Verified.}

To address the issues of the current \textit{SWE-Bench} datasets and ensure a more accurate evaluation of the models' effectiveness in resolving issues, we utilized a new dataset, \textit{SWE-Bench+}, which focuses on issues with no clear solution provided in the issue report and without potential risk of data leakage. 
The primary objective of \textit{SWE-Bench+} is to assess the models' ability to generate accurate patches for real-world GitHub issues without the risk of bias or prior exposure to the solution. %The dataset is based on issues and their corresponding pull requests created after the models' training cut-off dates to avoid any potential data leak, enabling a more reliable evaluation of the models' genuine capabilities in solving GitHub issues. 
To maintain consistency and fairness when comparing the resolution rates of the models on\textit{SWE-Bench} and \textit{SWE-Bench+}, we followed the same data collection methodology outlined in the SWE-Bench study, using their open-source scripts. 
%We enhanced the SWE-Bench dataset with more issues to address the problems we observed. We call this new dataset SWE-Bench+.

%\subsection{Creation of SWE-Bench+}
\begin{figure}[t!]
    \centering
    \includegraphics[width=0.75\textwidth]{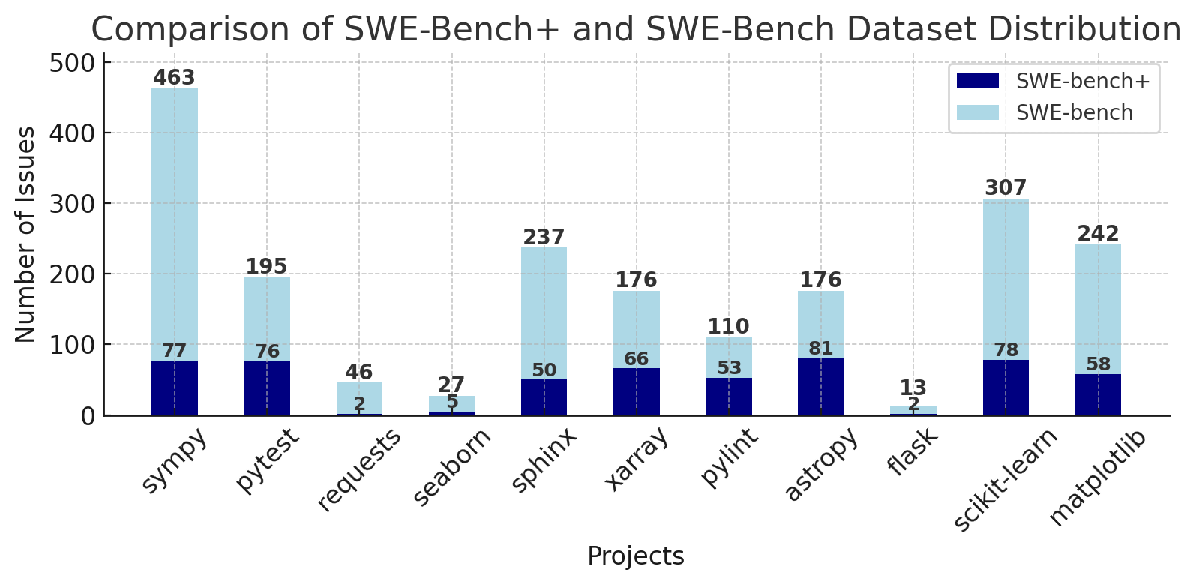} % Adjust width as needed
    \vspace{-0.1in}
    \caption{SWE-Bench+ dataset compared to SWE-Bench}
    \label{swe+_data}
\end{figure}
First, we selected the same 12 projects from \textit{SWE-Bench}, except Django, which was excluded as its issues are now tracked outside of GitHub. %The projects we focused on include sympy, matplotlib, scikit-learn, flask, astropy, requests, seaborn, sphinx, xarray, pylint, and pytest. 
Given that the LLMs we used in the models (GPT-4, GPT-3.5, and GPT-4o) were all trained on data up to October 2023, we collected issues that appeared after October 2023.  
Once the issues were collected, we applied the same filtering process described in the  \textit{SWE-Bench} study. This included using attribute filtering to retain only issues that resolve a problem and contribute tests, followed by an execution filter to keep only issues that install successfully and their PRs pass all tests. We then manually check all the instances to eliminate all the instances with a clear solution details in the issue report. By following this method, we obtained 548 task instances (issues) from the selected projects, with the distribution shown in Figure \ref{swe+_data}. 

% \song{Seems, we do not need this figure, it takes too much space, can you just mention how many from each project? If you really need it, reduce the height, also show the numbers of SWE-bench on each proj} 

%As outlined in \textit{SWE-Bench}, each task must be linked to the correct version of the project's repository at the time the issue occurred, ensuring that the appropriate installation instructions are applied. This versioning process is essential for enabling accurate execution-based evaluation. To get these versions, we follow the script provided by \textit{SWE-Bench} open-source code, specifically \textit{get\_versions.py} which ensures that in the version selected for every instance, the issue existed.

\section{Robustness of SWE-Bench+}
After collecting the data and identifying the correct repository versions, the next step was to run the selected models to generate patches that would address the issues. The models chosen for this task were \textit{SWE-RAG} with GPT-3.5 (turbo-16k-0613), \textit{SWE-RAG} with GPT-4 (1106), \textit{SWE-Agent} with GPT-4 (1106), and \textit{AutoCodeRover} (v20240620) paired with GPT-4o (2024-05-13). These models were selected based on their performance, large context window, cost, and the fact that they are open-source, making them more favorable compared to other options.

At the beginning of the project, \textit{SWE-Agent} had the highest resolution rate on the SWE-Bench leaderboard among all open-source models. Since we studied this model thoroughly in relation to SWE-Bench, we selected it to ensure a fair comparison. \textit{SWE-RAG+GPT-4} and \textit{SWE-RAG+GPT-3.5} were also chosen for the study due to their higher token limits compared to the \textit{Claude} models. Initial trials with \textit{SWE-Agent+Claude 3 Opus} and RAG-based Claude models (\textit{SWE-RAG+Claude 3 Opus} and \textit{SWE-RAG+Claude 2}) revealed that the token limits were exceeded before completing experiments on all 548 SWE-Bench+ instances. 
As the leaderboard frequently updates, with new models surpassing older ones, \textit{AutoCodeRover+GPT-4o} recently ranked among the top three models, and since it is open-source, we also included it in the study. We following their instructions to run the models on our \textit{SWE-Bench+} dataset.

%To execute the patch generation process, we used two primary scripts: the \texttt{run\_live.py} script for the RAG models on the SWE-Bench+ issue URLs, and the \texttt{run.py} script from the SWE-Bench open-source project for \textit{SWE-Agent+GPT-4}. Both scripts were run with the same configurations used in the original SWE-Bench setup. These scripts were responsible for generating the model-generated patches (\texttt{model\_patches}) intended to resolve the issues collected from the dataset.

Our evaluation followed a four-step technique, described in Figure~\ref{swe-image}, to ensure that the resolution rates of the models on SWE-Bench+ accurately reflected their correctness. Specifically, in \textbf{Step 1}, we stored all the model-generated diff files as patches in separate \textit{json} files, we then utilized the scripts provided by the SWE-Bench to evaluate the generated patches. 
% After running the scripts, we obtained evaluation logs, resolution rates, and lists of tests categorized under \textit{PASS\_TO\_PASS} and \textit{FAIL\_TO\_PASS}, along with their corresponding pass or fail statuses. 
In \textbf{Step 2}, once the evaluation results were generated, we manually reviewed the instances marked as \textit{resolved} and selected only those where all tests in \textit{PASS\_TO\_PASS} and \textit{FAIL\_TO\_PASS} had passed. In other words, we filtered out instances where the patches were successfully applied, and all associated tests passed.     
In \textbf{Step 3}, after filtering the resolved instances where all tests passed, we analyzed the resolution rates of the models. %The resulting resolution rates are: 4.20\% for \textit{SWE-RAG+GPT-4}, 3.65\% for \textit{SWE-RAG+GPT-3.5}, 5.84\% for \textit{SWE-Agent+GPT-4}, and 7.66\% for \textit{AutoCodeRover + GPT-4o}. These rates seemed unusually high compared to the results reported in Section 3, when only accounting for the "different implementation than gold patch" and "more comprehensive than gold patch" patterns. This discrepancy prompted further investigation into the resolved instances to identify patterns, as done in Section 3.
Finally, in \textbf{Step 4}, we performed a similar patch validation study as described in Section~\ref{sec:2}, comparing the model-generated patches to the gold patches. 
{We found that solution-leak related issues are now
resolved. However, a prominent issue with weak test cases persists. As shown in Table~\ref{swe-new-results},} 
on average, around 67.72\% of the resolved instances did not truly resolve the issue, despite passing all the tests. The prominent pattern identified was the models' inability to accurately locate the buggy files or lines. This raises concerns about the models' ability to locate the bug. In other cases, the model generated incomplete or incorrect fixes, not resolving the bug. These findings show that the issue with weak tests still persists and needs future investigation. 

As a result of the patch validation analysis, the resolution rates were 0.73\% for \textit{SWE-RAG+GPT-4}, 0.55\% for \textit{SWE-RAG+GPT-3.5}, 0.55\% for \textit{SWE-Agent+GPT-4}, and 3.83\% for \textit{AutoCodeRover+GPT-4o}. These rates are significantly lower than the reported resolution rates on the \textit{SWE-Bench} leaderboard were 1.31\% for \textit{SWE-RAG + GPT-4}, 0.17\% for \textit{SWE-RAG + GPT-3.5}, and 18.83\% for \textit{AutoCodeRover + GPT-4o}. This drop in performance for all models when moving from \textit{SWE-Bench Full} to \textit{SWE-Bench+}, highlights the impact of excluding suspicious fixes on the overall resolution rates.

% \song{add a new table to show the performance of these models on SWE-bench Full and your SWE-bench+}

   \begin{table}[t]
\centering
\caption{Performance of different models on \textit{SWE-bench+}}
\label{swe-new-results}

\resizebox{0.95\textwidth}{!}{%
\begin{tabular}{lrrrr}
\toprule
\textbf{Pattern} & \textbf{SWE-RAG+GPT-4} & \textbf{SWE-RAG+GPT-3.5} & \textbf{SWE-Agent+GPT-4} & \textbf{AutoCodeRover+GPT-4o} \\
\midrule
Correct                      & \textbf{4}   & 3   & 3   & \textbf{21} \\ \hline\hline
%\midrule
(Suspicious) Different files/functions changed & \textbf{16} & 11   & 14  & 7  \\    %  & \textbf{3}   & 2   & \textbf{3}   & 2  \\
(Suspicious) Incorrect fixes                   & 2   & 6   & 3   & \textbf{15} \\
(Suspicious) Incomplete fixes                 & 1   & 0   & 0   & \textbf{3}  \\

\midrule
\textbf{Total} & 23 & 20 & 32 & \textbf{42} \\
\bottomrule
\end{tabular}}

\end{table}

\section{Effectiveness-aware Evaluation}
\label{sec:cost}

\begin{table}[t!]
\centering
\caption{Average cost of different models on \textit{SWE-Bench+}}
\label{tab:cost}
\resizebox{0.95\textwidth}{!}{%
\begin{tabular}{l|c|c|c}
\hline
\textbf{Model}             & \textbf{Avg cost per instance} & \textbf{cost per issue fixing} &{\textbf{Avg time per instance}} \\ \hline
\textit{RAG+GPT 4}         & \$0.24                         & \$32.5  & 30 seconds                                           \\ \hline
\textit{RAG+GPT 3.5}       & \$0.05                         & \$10.0     & 30 seconds                                           \\ \hline
\textit{SWE\_AGENT+GPT 4}  & \$3.59                         & \$655.0       & 4 minutes                                        \\ \hline
\textit{AutoCode Rover+GPT 4o} & \$0.46                         & \$12.61       & 4.5 minutes                                      \\ \hline
\end{tabular}
}
\label{cost}

\end{table}

% \song{if we have time, just add a new column in table 3}
Despite the notable success in resolving issues, we also observed significant variations in the costs of the approaches tested. While some models excelled in accuracy and efficiency, they required more computational resources, longer processing times, and higher costs. Specifically, \textit{SWE-Agent+GPT-4} and \textit{AutoCodeRover+GPT-4} had the longest code generation times, with \textit{SWE-Agent+GPT-4} averaging around 4 minutes per instance, resulting in a total of approximately 37 hours to generate patches for SWE-Bench+ issues. The average generation time for \textit{AutoCodeRover} was 4.5 minutes per instance, resulting in 41 hours in total to generate patches for all SWE-Bench+ instances, making it a top performer in issue resolution for SWE-Bench+ instances. However, this disparity highlights the trade-offs between performance and cost-effectiveness, particularly for models like \textit{SWE-Agent+}, where balancing time, cost, and resource allocation is critical for real-world applications.

In terms of cost, we identified two metrics, i.e., the average cost per instance (calculated by dividing the total cost by the 548 instances in SWE-Bench+ that the models were tested on) and the effectiveness-aware cost per instance (calculated by dividing the total cost by the number of instances successfully resolved by the model). %The first component is calculated by dividing the total cost by the 548 instances in SWE-Bench+ that the models were tested on. The second component reflects the total cost divided by the number of instances successfully resolved by the model.
The detailed cost of each model measured by the two metrics is shown in Table~\ref{cost}. 
Overall, \textit{SWE-Agent+GPT-4} was the most expensive model, with an average cost of \$0.24 per instance and an effectiveness-aware cost of \$32.5 per issue fixed, as shown in Table \ref{cost}. Despite its high cost, its performance was comparable to \textit{RAG+GPT-4}, which was much more cost-efficient, with an average cost of \$0.05 per instance and an effectiveness-aware cost of \$10.0. On the other hand, \textit{AutoCodeRover+GPT-4} delivered the highest resolution rate of 3.83\% among all models. Although relatively costly on average, especially for larger datasets with an average cost of \$0.46 per instance, \textit{AutoCodeRover's} effectiveness-aware cost was relatively low at \$12.61 per issue fixed, given its high-resolution rate compared to the \textit{SWE-Agent} approach. %\textit{RAG+GPT-4} averaged \$0.24 per instance and an effectiveness-aware cost of \$10, making it significantly more cost-effective than \textit{SWE-Agent+GPT-4}, which had a total cost of \$655. 
Meanwhile, \textit{RAG+GPT-3.5} had the lowest average cost, at \$0.05 per instance, but a relatively high effectiveness-aware cost of \$32 due to its poorer performance in resolving issues correctly.

We recommend that future research not only assess the accuracy of the proposed models but also take into account the financial cost associated with their implementation and operation, ensuring they are not only performant but also practical for large-scale and long-term use.
% \song{explain your results, and comparisons results under different scenairos} \reem{what scenarios?}

%\song{you can motivate this a little bite: Despite the notable performance in resolving issues, we also observe a significant variation in the costs of the approaches tested.....While some methods excel in terms of accuracy and efficiency, they may require more computational resources, longer processing times, or xxx. Others, although less precise, may offer more cost-effective solutions in terms of time and resource allocation. This disparity highlights the trade-offs between performance and cost-effectiveness, underscoring the importance of considering both factors when selecting the most appropriate approach for real-world applications. }

%\song{analyze the financial of each experimented approach 1) each issue an average; 2) effectiveness-aware bug-fixing performance, average cost per bug fixed (total cost on the 548 issues / (number of fixed bugs))}

\section{Related Work}
\paragraph{LLM for Software Engineering.} 
%\song{I have updated the title to "LLM for SE", please rewrite this section according, not only bug fix}
%\song{read the follow papers, following their related work to write yours}
 % SWE-bench-Java: https://www.arxiv.org/pdf/2408.14354; 
%  https://arxiv.org/abs/2305.01210;  
 % https://arxiv.org/abs/2310.06770
% \todo{Haoran, you can work on this part}
Large Language Models (LLMs) have emerged as powerful tools and demonstrated impressive capabilities in various software engineering tasks, including code generation \cite{jiang2024surveylargelanguagemodels,10.1145/3650203.3663334, chen2021evaluatinglargelanguagemodels,
LUO2024100488,10.1145/3597503.3639219}, program repair \cite{zhang2024criticalreviewlargelanguage, 10.1145/3650212.3680328,DEFITERODOMINGUEZ2024109291} and bug detection \cite{alrashedy2024languagemodelsbetterbug, 10.1145/3660773}. The development of code generation benchmarks has been crucial for evaluating LLM performance. Notably, HumanEval \cite{chen2021evaluatinglargelanguagemodels} was introduced to assess the functional correctness of code generated by LLMs. Building on this foundation, AlphaCode \cite{Li_2022} demonstrated competitive performance in solving complex programming problems. To address limitations in existing benchmarks, EvalPlus \cite{liu2024your} enhanced HumanEval with more comprehensive test cases and revealed a significant overestimation of LLM performance in previous evaluations. LLMs also have shown promising results in program repair and bug detection. For example, AlphaRepair \cite{xia2022trainingrepairingpleaserevisiting} employed a zero-shot learning approach that outperformed traditional automated program repair (APR) tools. Further research demonstrated that LLMs could surpass existing APR techniques, particularly when fine-tuned on domain-specific data \cite{10.1109/ICSE48619.2023.00129}. The application of LLMs in bug detection with innovative approaches like FUzzGPT \cite{deng2023largelanguagemodelsedgecase} and TitanFuzz \cite{deng2023largelanguagemodelszeroshot} leveraging these models to generate edge-case test inputs and perform mutation-based fuzzing for deep learning libraries. 
There are several comprehensive studies have explored LLM applications across various software engineering domains \cite{fan2023largelanguagemodelssoftware, hou2024largelanguagemodelssoftware}, delved into the natural language to code generation \cite{zan2023largelanguagemodelsmeet}, and analyzed the evolution and performance of Code LLMs across different tasks \cite{zheng2024surveylargelanguagemodels}. %Recent innovations have also explored novel ways to leverage LLMs, such as CODET \cite{chen2022codetcodegenerationgenerated} that combined code generation with test case generation, and PanGu-Coder2 \cite{shen2023pangucoder2boostinglargelanguage} incorporated ranking feedback to enhance LLM performance in code generation. The potential of LLMs has even extended to hardware security, showing promise in preparing security-relevant bugs in hardware designs \cite{Ahmad_2024,ALSAQER2024}.

\paragraph{Benchmark Dataset Quality for Code Generation.} 
% \song{Add this paper}
% \cite{liu2024your}
To achieve accurate and reliable outcomes in code generation tasks with LLMs, it is essential to use high-quality evaluation benchmark datasets during the training and evaluation steps \cite{shi2022we,jimenez2024swebenchlanguagemodelsresolve}. With the growing interest and significance of code generation and program repair in software engineering, this highlights the critical need for trustworthy and reliable evaluation benchmark datasets \cite{jimenez2024swebenchlanguagemodelsresolve, zan2024swebenchjavagithubissueresolving}. To tackle this challenge, SWE-bench was developed to assist developers in evaluating LLMs using real-world GitHub issues~\cite{jimenez2024swebenchlanguagemodelsresolve}. %By reflecting the complex problems faced by developers, the dataset helps uncover potential limitations of LLMs in these areas \cite{jimenez2024swebenchlanguagemodelsresolve}. 
%\song{this is not related work}
%One of the key challenges in current state-of-the-art code generation benchmarks is the prevalence of noise and inconsistencies, which highly correlates to the performance of LLMs when evaluated against these datasets \cite{jimenez2024swebenchlanguagemodelsresolve}. This challenge underscores the need for higher-quality benchmark datasets and calls for increased research efforts to improve and refine existing benchmarks, ensuring their accuracy and reliability.
%Frameworks such as MAGIS and CodeR utilized advanced methods, including multi-agent techniques, to improve the quality of evaluation benchmarks. 
%These studies emphasize the critical role of task-specific data preparation techniques in generating more reliable and relevant datasets. 
A critical part of making a reliable dataset is to make it representative of real-world and complex software challenges and issues~\cite{chen2024coderissueresolvingmultiagent, tao2024magisllmbasedmultiagentframework}. 
%\song{I think I mention this paper https://www.eecs.yorku.ca/~wangsong/papers/fse22b.pdf and the related work in this paper }
% \reem{discuss AutoCodeRover how it uses tree of though for github resolution and 
%  how models trained on well-prepared datasets are better equipped to handle sophisticated tasks, such as autonomous program improvement and deep reasoning for issue resolution} [https://arxiv.org/abs/2404.05427] [https://arxiv.org/abs/2405.13057].
%Building upon the importance of high-quality benchmark datasets for code generation, AutocodeRover \cite{zhang2024autocoderoverautonomousprogramimprovement} shows how advanced techniques can leverage well-prepared datasets to tackle complex software engineering tasks. AutocodeRover utilizes a sophisticated reasoning framework akin to the Tree of Thought (ToT) approach, employing iterative context retrieval and patch generation to resolve GitHub issues autonomously\cite{larosa2024githubissuessolvedtree, zhang2024autocoderoverautonomousprogramimprovement}. Its method of stratified context search enables it to gather relevant code snippets progressively, enhancing its understanding of intricate issue resolution\cite{zhang2024autocoderoverautonomousprogramimprovement}. 
% \reem{discuss need for diversity and comprehensiveness in "Diversity Empowers Intelligence" where diverse data enhance the performance of LLMs in code generation tasks }
 % [https://arxiv.org/abs/2408.07060]. 
The ``Diversity Empowers Intelligence'' (DEI) framework \cite{zhang2024diversityempowersintelligenceintegrating} further shows how diversity and comprehensiveness of data can enhance the performance of Large Language Models (LLMs) in code generation tasks. %By leveraging diverse SWE agents, DEI capitalizes on the complementary strengths of these agents in software engineering tasks such as bug fixing and patch generation by allowing DEI to harness a broader spectrum of expertise, which improves the collective performance significantly \cite{zhang2024diversityempowersintelligenceintegrating}. 
Given the critical need for high-quality benchmark datasets in evaluating code generation models, Liu et al. \cite{liu2024your} introduce EvalPlus, a framework specifically designed to improve the evaluation of LLM-generated code. This framework addresses challenges like insufficient tests and noisy benchmarks by augmenting existing datasets with automated test input generation using LLM and mutation-based strategies. Their findings show the need for more rigorous testing in LLM code generation to improve benchmark quality in the field.
 % \reem{discuss "LLM Agent Coordination for Improved GitHub Issue Resolution" to show that high quality and diverse data can lead to more effective collaborative problem-solving among LLMs}

%\song{discuss the difference of your work from the above ones}
\section{Conclusion}
In this paper, we presented the first empirical study on the robustness of the \textit{SWE-Bench} dataset.  
Our study identified significant limitations in the original \textit{SWE-Bench} dataset, particularly issues with solution leakage and weak test cases, which undermined the reliability of previous model assessments. %Specifically, we found that 32.67\% of the issues resolved by LLMs involved answer leakage, and 31.08\% passed despite weak or incomplete tests. These critical flaws lowered the actual performance of LLMs such as SWE-Agent + GPT-4, whose accuracy dropped from 12.47\% to 3.97\% when the suspicious data were excluded. 
To address these challenges, we introduced SWE-Bench+, a dataset free from solution leakage and built with issues created after LLM training cut-off dates to ensure more rigorous and accurate evaluations. Through extensive testing, we demonstrated that while \textit{SWE-Bench+} resolves the data leakage concerns, weak test cases continue to pose challenges, with model resolution rates dropping further in this refined environment. %For example, SWE-Agent + GPT-4's performance fell to 0.55\%, underscoring the difficulty in resolving complex real-world issues without access to prior solutions. 
Despite the reduced pass rates, \textit{SWE-Bench+} establishes a more reliable framework for assessing the true capabilities of LLMs in software development, offering insights into how these models can be better developed and evaluated. Future work should focus on improving the test case robustness in SWE-Bench+ and exploring more effective strategies for filtering data to further minimize biases and inaccuracies.

%\textcolor{blue}{
%By refining the evaluation methodologies and datasets, this study contributes to advancing the reliability and performance of LLMs in software engineering tasks. As LLMs continue to evolve, benchmarking frameworks like SWE-Bench+ will play a crucial role in ensuring that their practical capabilities are accurately assessed and that their deployment in real-world software development tasks is both effective and trustworthy.}
%A limitation of our study is that we did not evaluate all available models, leaving the performance of newer, top-performing models still uncertain. As the SWE-Bench leaderboard continues to evolve, newer models are surpassing previous ones in issue resolution rates. While the issue of solution leakage remains a concern, it is possible that the correct resolution rate is higher for instances where the solution is not provided in the issue report.

For future work, further investigation into the issue of weak test cases is needed, along with suggestions for improving test quality to create more accurate and relevant test suites. Another potential avenue of research could explore the underlying causes of the high failure rates and propose strategies to mitigate them. Additionally, similar studies could be conducted on other leading evaluation benchmarks, such as Human-Eval, to compare results and identify broader patterns.

\newpage

\label{others}

\bibliography{reference}
\bibliographystyle{iclr2024_conference}

\end{document}